\documentclass[twocolumn,amsmath,amssymb,color,superscriptaddress]{revtex4}
\usepackage{graphicx}
\begin{document}

\title{New carbon allotropes in $sp+sp^3$ bonding networks consisting of C$_8$ cubes }

\author{Jian-Tao Wang}
\email[e-mail address:]{wjt@aphy.iphy.ac.cn}
\affiliation{Beijing National Laboratory for Condensed Matter Physics,
Institute of Physics, Chinese Academy of Sciences, Beijing 100190, China}
\affiliation{School of Physics, University of Chinese Academy of Sciences, Beijing 100049, China }

\author{Changfeng Chen}
\affiliation{Department of Physics and High Pressure Science and
Engineering Center, University of Nevada, Las Vegas, Nevada 89154, USA}
\author{Hiroshi Mizuseki}
\affiliation{Computational Science Research Center, Korea Institute of Science and Technology (KIST), Seoul 02792, Republic of Korea}

\author{Yoshiyuki Kawazoe}
\affiliation{New Industry Creation Hatchery Center, Tohoku University, Sendai 980-8579, Japan}
\affiliation{Department of Physics and Nanotechnology, SRM University, Kattankulathur, 603203, TN, India}

\date{\today}

\begin{abstract}
We identify by {\it ab initio} calculations a new type of three-dimensional carbon allotropes constructed by inserting acetylenic or diacetylenic bonds into a body-centered cubic C$_8$ lattice. The resulting $sp+sp^3$-hybridized cubane-yne and cubane-diyne structures consisting of C$_8$ cubes can be characterized as a cubic crystalline modification of linear carbon chains, but energetically more favorable than the simplest linear carbyne chain and the cubic tetrahedral diamond and yne-diamond consisting of C$_4$ tetrahedrons. Electronic band calculations indicate that these new carbon allotropes are semiconductors with an indirect band gap of 3.08 eV for cubane-yne and 2.53 eV for cubane-diyne. The present results establish a new type of carbon phases consisting of C$_8$ cubes and offer insights into their outstanding structural and electronic properties.
\end{abstract}
\maketitle

\section{Introduction}
The valence electrons of carbon atom are capable of forming $sp^3$-, $sp^2$- and $sp$-hybridized states that possess a wide range of properties with numerous applications in many areas of science and technology \cite{ct1,sp123,bat1989}. The two naturally occurring crystalline carbon structures in pure elemental form, i.e., graphite and diamond, possess all $sp^2$ or $sp^3$ carbon-carbon bonds, respectively. During the past thirty years, considerable theoretical and experimental efforts have been made to assess new potential carbon allotropes \cite{jap2010,hdg2005,sc48,p1,ST12,mao2003,mcarbon,wcarbon,wcarbon2,wcarbon3,zcarbon,note5,science,BC12,sc24,nl2015,km2015,wen2016,wjt2016prl,lzz2016,c60,ct,pha,cg1992}.
These efforts have led to the discovery of a large variety of carbon allotropes with remarkable properties; most intriguing among them are the zero-dimensional fullerenes \cite{c60}, one-dimensional carbon nanotubes \cite{ct}, two-dimensional graphene \cite{pha}, and cubic polybenzene \cite{cg1992}. Recently, linear carbyne, the simplest one-dimensional (1D) carbon chain has been synthesized \cite{nc2010} in alternating single and triple carbon-carbon bonding state. Meanwhile, the so-called graphyne \cite{jcp1987rh} and graphdiyne \cite{acie1997} are proposed by replacing one-third of the C-C bonds in graphene sheet with acetylenic ($-$C$\equiv$C$-$) or diacetylenic ($-$C$\equiv$C$-$C$\equiv$C$-$) linkages.
Experimentally, large-scale graphyne and graphdiyne films composed of $sp+sp^2$ hybrid network have been successfully synthesized \cite{cc2010li,pssc2013al,ol2000jm}. Inspired by two-dimensional $sp+sp^2$ hybridized carbon allotropes, three-dimensional (3D) polybenzene-ynes in $sp+sp^2$ hybridized bonds with phenylic rings and acetylenic chains have been reported recently \cite{wjt2016sr} by inserting acetylenic or diacetylenic bonds into an all $sp^2$-hybridized rhombohedral polybenzene lattice \cite{ct2}. Moreover, 3D $sp$+$sp^3$-yne-diamond was suggested by inserting acetylenic linkers into all the carbon-carbon bonds in cubic diamond \cite{nmat868,prb2012jy,jmc2013lh}. Meanwhile, a tetrayne-carbon (TY-carbon) \cite{prb2012jy} has been designed by inserting yne bonds into the cubic tetrahedral diamond (T-carbon) \cite{Tcarbon}. Such porous yne-diamond carbon framework would be advantageous for gas storage \cite{jmc2013lh}. However, these $sp$+$sp^3$ hybrid network structures are energetically less stable than the simplest 1D linear carbon chains \cite{wjt2016sr}.

Beside cubic tetrahedral diamond, a supercubane consisting of a body-centered cubic array of C$_8$ cubes was theoretically predicted \cite{jacs1985} and synthesized recently by using a pulsed-laser induced liquid-solid interface reaction \cite{bctc8}. To understand the stability of yne-diamond-like carbon framework structures, we have performed a detailed {\it ab initio} study on a new type of 3D carbon allotropes constructed by inserting acetylenic or diacetylenic bonds into the supercubane lattice \cite{jacs1985}.
The resulting all carbon cubane-yne and cubane-diyne network structures can be characterized as a cubic crystalline modification of linear carbon chains, but they are energetically more favorable than the carbyne chain and the cubic tetrahedral diamond and yne-diamond consisting of C$_4$ tetrahedrons. Simulated phonon spectra reveal that these structures are
dynamically stable. Electronic band calculations indicate that they are semiconductors with an indirect band gap of 3.08 eV for cubane-yne and 2.53 eV for cubane-diyne.

\section{COMPUTATIONAL METHOD}
Our calculations were carried out using the density functional theory as implemented in the Vienna {\it ab initio} simulation package (VASP) \cite{vasp}. The generalized gradient
approximation (GGA) developed by Armiento-Mattsson (AM05) \cite{am05} were adopted for the exchange-correlation potential. The all-electron projector augmented wave (PAW)
method \cite{paw} was adopted with 2$s^2$2$p^2$ treated as valence electrons. A plane-wave basis set with a large energy cutoff of 800 eV was used. Convergence criteria employed for both the electronic self-consistent relaxation and the ionic relaxation were set to 10$^{-8}$ eV and 0.01 eV/\AA\ for energy and force, respectively. A hybrid density functional method based on the Heyd-Scuseria-Ernzerhof scheme (HSE06) \cite{HSE06} was used to calculate electronic properties. Phonon calculations were based on the supercell approach \cite{pho1} using the phonopy code \cite{pho2}.

\begin{figure}[b]
\includegraphics[width=8.00cm]{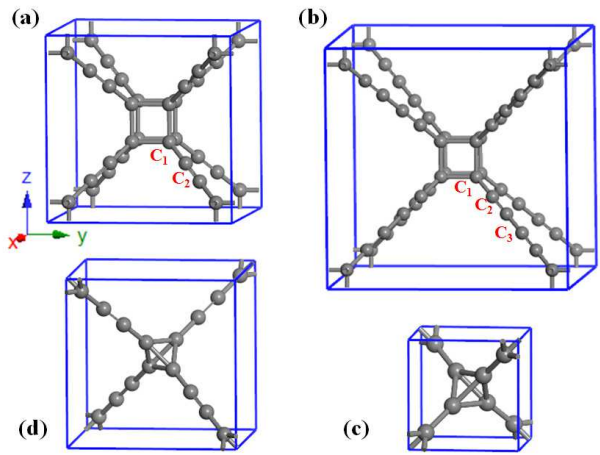}
\caption{Cubic yne-carbon structures consisting of C$_8$ cubes or C$_4$ tetrahedrons. (a) Cubane-yne in $sp+sp^3$-hybridized bonding network with lattice parameter a = 7.8370 \AA. The 16$f_1$ (0.1014, 0.1014, 0.1014) atoms form two C$_8$ cubes and 16$f_2$ (0.2051, 0.2051, 0.2051) atoms form eight acetylenic yne-bonds located between the C$_8$ cubes;
(b) Cubane-diyne in $sp+sp^3$-hybridized bonding network with lattice parameter a = 10.8108 \AA.  The atoms on 16$f_1$ (0.0737, 0.0737, 0.0737) site form two C$_8$ cubes. The atoms on 16$f_2$ (0.1485, 0.1485, 0.1485) and 16$f_3$ (0.2141, 0.2141, 0.2141) sites form eight diyne-bonds located between the C$_8$ cubes; (c) T$_4$-carbon consisting of a simple cubic array of C$_4$ tetrahedrons in $sp^3$ bonding state; (d) T$_4$-yne consisting of C$_4$ tetrahedrons and acetylenic yne-bonds in $sp+sp^3$-hybridized bonding network.}
\end{figure}

\section{RESULTS AND DISCUSSION}
The simplest cubane-yne network can be constructed by inserting acetylenic ($-$C$\equiv$C$-$) yne-bonds into the supercubane lattice [see Fig. 1(a)]. This new carbon phase has an $Im\bar{3}m$ ($D_{O}^9$) symmetry, the same as that of supercubane lattice \cite{jacs1985}. In the body centered cubic representation, it has a 32-atom unit cell with lattice parameter
$a$ = 7.8370 \AA, occupying the 16$f_1$ (0.1014, 0.1014, 0.1014) and 16$f_2$ (0.2051, 0.2051, 0.2051) Wyckoff positions, denoted by C$_1$ and C$_2$, respectively. The carbon atoms on the C$_1$ positions form two C$_8$ cubes, as in supercubane with $sp^3$ hybridization, while the carbon atoms on the C$_2$ positions form eight triple yne-bonds located between the C$_8$ cubes with $sp$-hybridization. There are three distinct carbon-carbon bond lengths, a longer bond of 1.589 \AA\ (C$_1$-C$_1$) in the cubes and two shorter bonds of 1.408 \AA\ (C$_1$-C$_2$) and 1.220 \AA\ (C$_2$-C$_2$) associated with the single and triple bond in carbyne chains, respectively. Meanwhile, it has three different bond angles, 180$^o$ for $\angle$C$_1$-C$_2$-C$_2$ along the carbyne chains, 90$^o$ for $\angle$C$_1$-C$_1$-C$_1$ inside and 125.26$^o$ for $\angle$C$_1$-C$_1$-C$_2$ outside the cubes.

Figure 1(b) shows the cubane-diyne network by inserting diacetylenic ($-$C$\equiv$C$-$C$\equiv$C$-$) bonds between the cubes in the supercubane lattice.  The resulting structure has a 48-atom cubic unit cell with an equilibrium lattice parameter $a$ = 10.8108 \AA, occupying the 16$f_1$ (0.0737, 0.0737, 0.0737), 16$f_2$ (0.1485, 0.1485, 0.1485) and 16$f_3$ (0.2141, 0.2141, 0.2141) Wyckoff positions, denoted by C$_1$, C$_2$, and C$_3$, respectively. The carbon atoms on the C$_1$ positions form two C$_8$ cubes, while the carbon atoms on the C$_2$ and C$_3$ positions form eight diyne-bonds located between the C$_8$ cubes. There are four distinct carbon-carbon bond lengths, a longer bond of 1.593 \AA\ (C$_1$-C$_1$) is associated with the carbon atoms in the cubic units and three shorter bonds of 1.401 \AA\ (C$_1$-C$_2$), 1.228 \AA\ (C$_2$-C$_3$), and 1.346 \AA\ (C$_3$-C$_3$) are along the chains between C$_8$ cubes. Meanwhile, as in cubane-yne, there are three different bond angles, 180$^o$ for $\angle$C$_1$-C$_2$-C$_3$ along the carbyne chains, 90$^o$ for $\angle$C$_1$-C$_1$-C$_1$ inside and 125.26$^o$ for $\angle$C$_1$-C$_1$-C$_2$ out of the cubes. It is noted that these 3D carbon network structures are topologically corresponding to the three-fold carbon chains with bond change from triple to single bond in the cubes, and thus cubane-yne and cubane-diyne can be regarded as the cubic crystalline modification of carbon chains.

\begin{table*}
\caption{Calculated equilibrium structural parameters (space group, volume $V_0$, lattice parameters $a$ and $c$, bond lengths $d_{C-C}$), total energy $E_{tot}$, bulk modulus $B_0$, and electronic band gap $E_g$ for diamond, graphite, supercubane, cubane-yne, and cubane-diyne at zero pressure, compared to available experimental data \cite{data1}.}
\begin{tabular}{lllcccccc} \hline
\hline
Structure &  Method & $V_0$(\AA$^3$/atom)  & a (\AA) & c (\AA) & $d_{C-C}$ (\AA) & $E_{tot}$ (eV) &   $B_0$ (GPa)   & $E_g$ (eV)\\
\hline
Diamond  ($Fd\bar{3}m$)  &  AM05 \cite{wjt2016sr}         & 5.604 & 3.552 & & 1.538                     &  -9.018 &  451   & 5.36 \\
                      & Exp \cite{data1}  & 5.673 & 3.567 & & 1.544     &                         &  446   & 5.47  \\

Supercubane ($Im\bar{3}m$)  &  AM05     & 7.148  & 4.853 &    & 1.470, 1.578  & -8.355 &  329   & 4.17 \\

Cubane-yne  ($Im\bar{3}m$)  &  AM05     & 15.052  & 7.837 &    & 1.220$-$1.589  & -8.100 &  148   & 3.08 \\

Cubane-diyne ($Im\bar{3}m$) &  AM05     & 26.345  & 10.811 &    & 1.228$-$1.593  & -8.048 &  84.6   & 2.53 \\

Graphite  ($P6_3/mmc$)  &  AM05 \cite{wjt2016sr}  & 8.813  & 2.462 & 6.710    & 1.422                      &  -9.045 &  280   &  \\
             & Exp \cite{data1} & 8.783  & 2.460 & 6.704    & 1.420                      &         &  286   &   \\
\hline
\hline
\end{tabular}
\end{table*}

\begin{figure}[b]
\includegraphics[width=8.20cm]{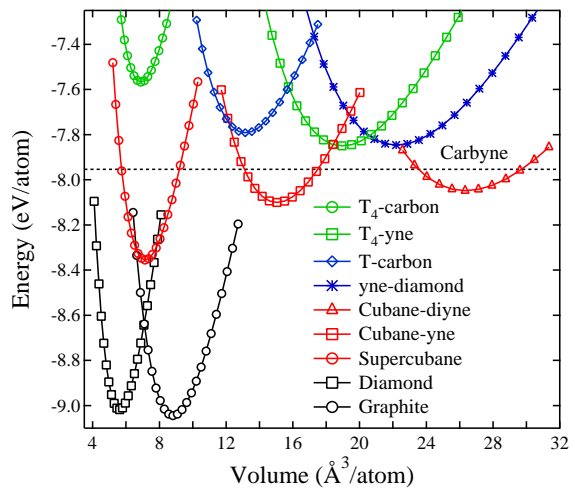}
\caption{The total energy as a function of volume per atom for cubic supercubane, cubane-yne, and cubane-diyne compared to those of T$_4$-carbon, T$_4$-yne, T-carbon \cite{Tcarbon}, yne-diamond \cite{prb2012jy}, graphite and diamond. The dashed line indicates the energy level of linear carbyne chain.}
\end{figure}

Figure 2 shows the calculated total energy versus the volume per atom for supercubane, cubane-yne, and cubane-diyne in comparison with the results for diamond, graphite, carbyne, yne-diamond \cite{prb2012jy}, and T-carbon \cite{Tcarbon}. The energetic data establish the following stability sequence:  T-carbon $<$ yne-diamond $<$ carbyne $<$ cubane-diyne $<$ cubane-yne $<$ supercubane. It is seen that both cubane-yne and cubane-diyne are located between the energy range for supercubane and carbyne, and energetically more stable than the 1D carbyne, while yne-diamond and T-carbon are less stable than carbyne with an energy loss of 0.11 eV and 0.16 eV per atom, respectively.

To better understand the energetic stability of yne-diamond, we further introduce a new simple cubic T$_4$-carbon structure [see Fig. 1(c)] (the 2$\times$2$\times$2 supercell containing two interwinding T-carbon networks) and study its tetrayne (T$_4$-yne) structure [see Fig. 1(d)]. It is shown that T$_4$-yne carbon is more stable than the cubic T$_4$-carbon and T-carbon, as stable as yne-diamond \cite{prb2012jy}, but less stable than the linear carbon chains (see Fig. 2). These results suggest that yne-diamond like T$_4$-yne carbon consisting of C$_4$ tetrahedrons is less stable than carbyne due to the less favorable tetrahedral units in T$_4$-carbon and T-carbon \cite{Tcarbon}.
The calculated equilibrium structural parameters, total energy, and bulk modulus for diamond, supercubane, cubane-yne, and cubane-diyne, and graphite are listed in Table I and compared to available experimental data \cite{data1}.

\begin{figure}
\includegraphics[width=8.00cm]{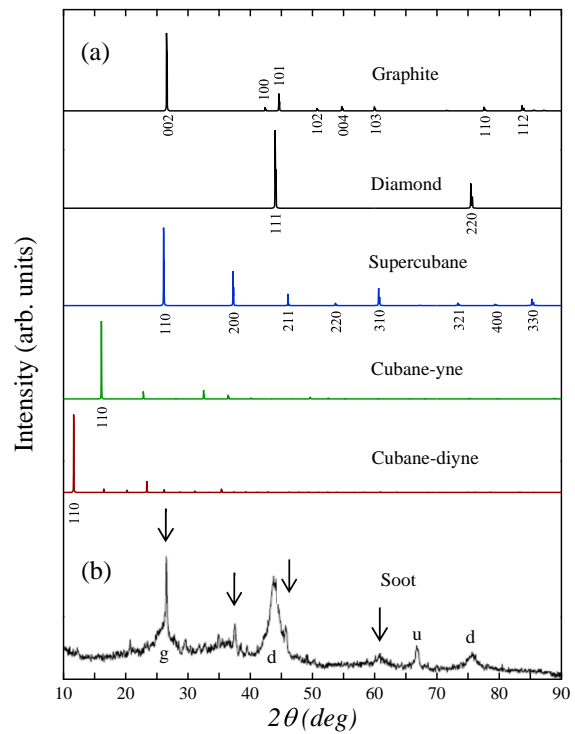}
\caption{Comparison of simulated and experimental X-ray diffraction (XRD) patterns. (a) Simulated XRD patterns for graphite, diamond,  supercubane, cubane-yne, and cubane-diyne.
(b) Experimental XRD patterns for detonation soot \cite{Chemosphere65}. $g$, $d$, and $u$ indicate graphite, diamond, and unknown-carbon, respectively. The X-ray wavelength is 1.54059 {\AA}.}
\end{figure}

To further establish the link between the calculated results and experiment, we plot the simulated x-ray diffraction (XRD) spectra of graphite, diamond, supercubane, cubane-yne,
and cubane-diyne in Fig. 3, compared to the experimental data for detonation soot \cite{Chemosphere65}. In the experimental XRD data \cite{Chemosphere65}, the most prominent peaks arise from graphite ($g$) and diamond ($d$); the peaks for supercubane at (110), (200), (211), and (310) match well with the experimental XRD spectra located at 25.9$^{\circ}$, 37.4$^{\circ}$, 46.0$^{\circ}$, and 60.3$^{\circ}$, respectively. It is noted that the main (101) peak at 25.9$^{\circ}$ for supercubane is very close to the main (002) peak at 26.5$^{\circ}$ for graphite, thus supercubane may coexist with graphite and diamond in the detonation soot \cite{Chemosphere65}. Meanwhile, the main (110) peak at 25.9$^{\circ}$ for supercubane should shift to 15.9$^{\circ}$ for cubane-yne and 11.6$^{\circ}$ for cubane-diyne carbon.

\begin{figure}
\includegraphics[width=8.0cm]{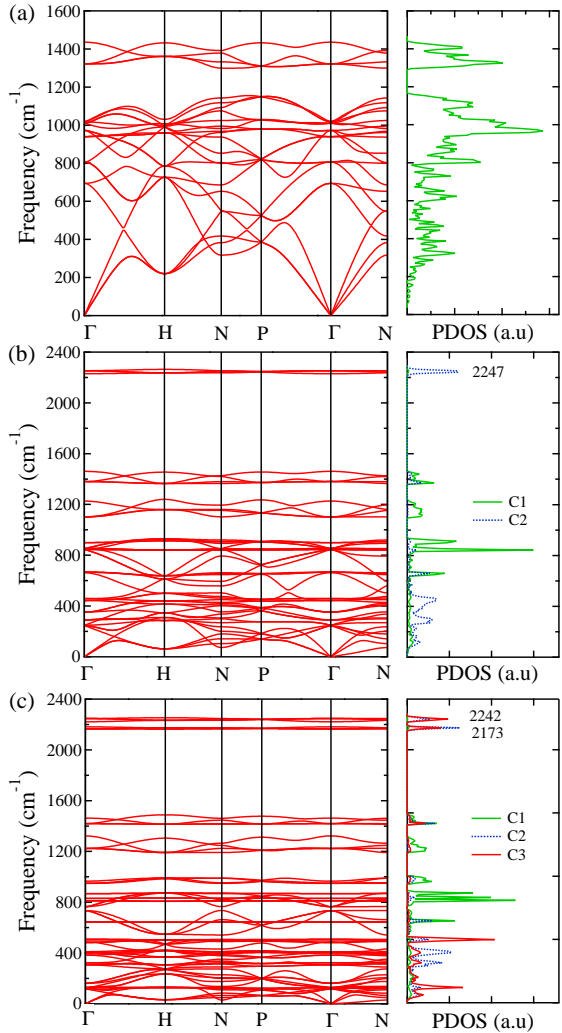}
\caption{Phonon band structures and density of states (PDOS) for supercubane (a) in all $sp^3$ bonds, cubane-yne (b) and cubane-diyne (c) in $sp$+$sp^3$ bonds.
The spectra due to the triple bonds occur around 2173$^{-1}$ and 2242 cm$^{-1}$.}
\end{figure}

\begin{figure}
\includegraphics[width=8.00cm]{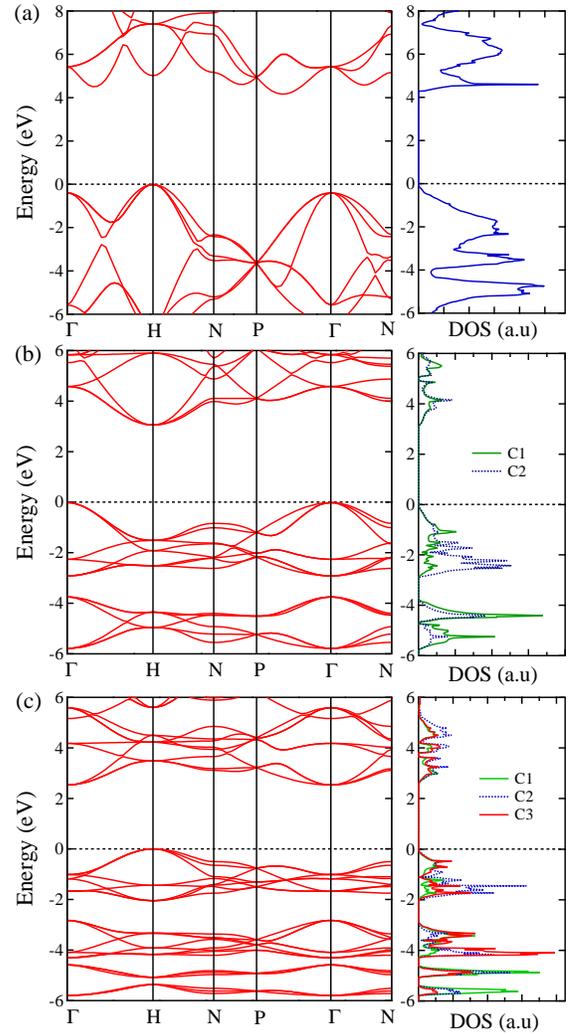}
\caption{Electronic band structures and density of states (DOS) for supercubane (a) in all $sp^3$ bonds, cubane-yne (b) and cubane-diyne (c) in $sp$+$sp^3$ bonds.
The Fermi level is set at zero eV as indicated by the dashed lines.}
\end{figure}

We next examine the dynamical stability of supercubane, cubane-yne, and cubane-diyne by phonon mode analysis. Figure 4(a) shows the phonon band structures and phonon density of states (PDOS) for supercubane in all $sp^3$ bonds. The highest phonon frequency is located at $\Gamma$ point with a value of $\sim$1435 cm$^{-1}$ due to the shorter bond length of 1.470 \AA\ between the C$_8$ cubes, which is higher than $\sim$1350 cm$^{-1}$ for perfectly $sp^{3}$ bonded diamond \cite{note15}, but closer to $\sim$1480 cm$^{-1}$ for SC48 carbon \cite{sc48}. Throughout the entire Brillouin zone, no imaginary frequencies are observed, confirming the dynamical stability of supercubane. Figure 4(b) shows the phonon band structures and PDOS for cubane-yne in $sp$+$sp^3$ bonding network. There is a large phonon band gap in the frequency range of 1462 and 2230 cm$^{-1}$. The vibrational modes due to the triple yne-bonds (C$_2$-C$_2$) can be observed clearly around 2247 cm$^{-1}$, while the vibrational modes due to the $sp^3$ bonds (C$_1$-C$_1$) are distributed below 1462 cm$^{-1}$. No imaginary frequencies were observed throughout the entire phonon band structures, thus confirming the dynamical stability of cubane-yne. Similar dynamical stability and vibrational modes are also confirmed for cubane-diyne as shown in Fig. 4(c). However, in the latter case there are two yne-modes around 2173 and 2242 cm$^{-1}$ related to the C$_2$ and C$_3$ carbon atoms. These vibrational modes arising from the triple yne-bonds are very close to experimental data 2189.8 cm$^{-1}$ and 1926.2 cm$^{-1}$ found in graphdiyne film \cite{cc2010li}.

Finally we discuss the electronic properties. The electronic band structures and density of states (DOS) are calculated based on the hybrid density functional method (HSE06) \cite{HSE06}. For supercubane, as shown in Fig. 5(a), the conduction band minimum is located along the P-$\Gamma$ direction and valence band maximum is located at the H point, showing a semiconducting behavior with an indirect band gap of 4.17 eV, smaller than 5.36 eV for diamond (see Table I). For cubane-yne, as shown in Fig. 5(b), the conduction band minimum and valence band maximum are located at the H and $\Gamma$ point, respectively, showing a semiconductor character with an indirect band gap of 3.08 eV, which is smaller than 4.17 eV for supercubane. Meanwhile, for cubane-diyne, as shown in Fig. 5(c), the conduction band minimum and valence band maximum are located at the $\Gamma$ and H point, respectively, also showing a semiconductor character but with a smaller indirect band gap of 2.53 eV, which is close to 2.56 eV found in the linear carbon chain \cite{nc2010}.
We can see that change of the carbon chain length has a direct and considerable influence on the electronic band gap.

\section{CONCLUSION}
In conclusion, we have identified by {\it ab initio} calculations a new type of three-dimensional carbon allotropes constructed by inserting acetylenic or diacetylenic bonds into a body-centered cubic supercubane lattice. The resulting all carbon cubane-yne and cubane-diyne network structures in $sp+sp^3$ bonding are topologically corresponding to the crystalline modification of linear carbon chains, and they are energetically more favorable than the $sp$-hybridized carbyne chains and the recently reported $sp+sp^3$-hybridized yne-diamond.  Phonon calculations show that these newly predicted structures are all dynamically stable.  Electronic band and density of states calculations indicate that both cubane-yne and cubane-diyne with yne-bonds are semiconductors with an indirect band gap of 3.08 eV and 2.53 eV, respectively. Our findings suggest a novel strategy in constructing carbon framework structures of yne-diamond and offer insights into their outstanding structural and electronic properties in $sp+sp^3$ bonding networks.

\section*{ACKNOWLEDGMENTS}
This study was supported by the National Natural Science Foundation of China (Grants No. 11674364) and the Strategic Priority Research Program of the Chinese Academy of Sciences (Grant No. XDB07000000). C.F.C. acknowledges support by DOE under Cooperative Agreement No. DE-NA0001982. H.M. is grateful for the financial support from the Korea Institute of Science and Technology (Grant No. 2Z04882 and  No. 2E26940).

\end{document}